\documentclass[%
 reprint,
 amsmath,amssymb,
 aps,
 prl,
floatfix,
]{revtex4-2}

\usepackage{comment}
\usepackage{graphicx}
\usepackage{dcolumn}
\usepackage{bm}
\usepackage{hyperref}
\usepackage{orcidlink}
\DeclareUnicodeCharacter{2009}{\,}

\begin{document}

\preprint{APS/123-QED}

\title{Hierarchical Three-Body Problem at High Eccentricities = Simple Pendulum}

\author{Ygal Y. Klein\orcidlink{0009-0004-1914-5821}}
\email{ygalklein@gmail.com}
 
\author{Boaz Katz\orcidlink{0000-0003-0584-2920}}
 
\affiliation{
 Dept. of Particle Phys. \& Astrophys., Weizmann Institute of Science,
 Rehovot 76100, Israel
}

\date{\today}

\begin{abstract}

 The gradual evolution of the restricted hierarchical three body problem is analyzed analytically, focusing on conditions of Kozai-Lidov Cycles that may lead to orbital flips from prograde to retrograde motion due to the octupole (third order) term which are associated with extremely high eccentricities. We revisit the approach described by Katz, Dong and Malhotra (\href{https://doi.org/10.1103/PhysRevLett.107.181101}{Phys. Rev. Lett. 107, 181101 (2011)}) and show that for most initial conditions, to an excellent approximation, the analytic derivation can be greatly simplified and reduces to a simple pendulum model allowing an explicit flip criterion. The resulting flip criterion is much simpler than the previous one but the latter is still needed in a small fraction of phase space. We identify a logical error in the earlier derivation but clarify why it does not affect the final results.
 
\end{abstract}

\maketitle
\paragraph{Introduction}
The dynamics of the restricted, hierarchical three-body problem (a test particle orbiting a central mass on a Keplerian orbit which is perturbed by a distant mass) involves oscillations of eccentricity and inclination on a timescale much longer than the orbital periods. Expanding the perturbing potential up to leading order in the small parameter of the ratio of semi major axes (quadrupole order) results in periodic oscillations which have been solved analytically (Kozai-Lidov Cycles, KLCs) \cite{kozai1962,lidov1962}\footnote{See recent historical overview including earlier relevant work by \citet{vonZeipel1910} in \citet{ito_2019}.}. The octupole term allows for the generation of extremely high eccentricities of the inner orbit and the possibility of a \textit{"flip"}, i.e a change from prograde to retrograde orbits, for perturbers on an eccentric orbit (Eccentric Kozai Lidov, EKL) \cite{katz2011,naoz2011,lithwick2011,naoz2013,Ford2000,li2014b,lei2022a,lei2022b} (for a review see \cite{naoz2016}). Very close approaches of the inner binary members due to the EKL have been argued to play a major role in a wide range of astrophysical phenomena, including satellites, planets, and black hole mergers \citep{naoz2012,teyssandier2013,stephan2016,liu2018,Angelo2022,Melchor2024,petrovich2015,stephan21}.

An analytical approximation for the EKL was derived in \cite{katz2011} by averaging the secular equations of motion over KLCs obtaining effective equations for the evolution of slow variables allowing for an analytical flip criterion to be derived. Recently, \cite{Sidorenko2018} and \cite{lei2022b} studied analytically the EKL dynamics through perturbative methods and mentioned a \textit{"pendulum"}-like structure in the resulting maps. Additionally, \cite{weldon2024} constructed an analytic approach for the secular descent and the timescale of the octupole term effect.

In this Letter the analysis of \cite{katz2011} is revisited. It is shown that for most initial conditions, to an excellent approximation, the effective equations reduce to those of a simple pendulum clarifying the dynamics and allowing the derivation of a flip criterion which is much simpler than that of \cite{katz2011}. For a small region of phase space the complex analytical criterion in \cite{katz2011} gives better reconstructions of the numerical results. A logical error in the derivation of \cite{katz2011} is identified and resolved.

\paragraph{Coordinate System}

Consider a test particle orbiting a central mass $M$ on an \textit{inner} orbit with semimajor axis $a$ and eccentricity $e$ and a distant mass $m_{per}$ on an \textit{outer} orbit with $a_{per},e_{per}$ where $a/a_{per}\ll1$. Following \cite{katz2011} we align the \textit{z} axis along the direction of the total angular momentum vector (which is the angular momentum vector of the outer orbit since the inner orbit is of a test particle). The \textit{x} axis is directed towards the pericenter of the (constant) outer orbit.
The dynamics of the test particle can be parameterized
by two dimensionless orthogonal vectors $\mathbf{j}=\mathbf{J}/\sqrt{GMa}$, where $\mathbf{J}$ is the specific angular momentum vector, and
$\mathbf{e}$ a vector pointing in the direction of the pericenter
with magnitude $e$. Using the variables $i_e$ and $\Omega_e$ as in \cite{katz2011} the eccentricity vector $\mathbf{e}$ is given by $\mathbf{e}=e\left(\sin i_e \cos \Omega_e,\sin i_e \sin\Omega_e,\cos i_e\right)$. We note that the usage of $\Omega_e$ as one of the canonical variables was suggested in \cite{Tremaine2001} (denoted $\theta_a$ therein) and the connection between $\Omega_e$ and the critical argument studied in \cite{lei2022a} ($\sigma_1$ therein) is discussed in the appendix of \cite{lei2022a}.

Expanding the perturbing potential to the third order term in the small parameter $a/a_{per}$ (the octupole term) and averaging over the two orbits results with the following potential (e.g, \citep{katz2011,liu2014,petrovich2015,tremaine2023,luo2016}):
\begin{equation}
\Phi_{\text{per}}=\Phi_{0}\left(\phi_{\text{quad}}+\epsilon_{\text{oct}}\phi_{\text{oct}}\right)\label{eq:phi_Per}
\end{equation}
where 
\begin{equation}
  \phi_{\text{quad}}=\frac{3}{4}\left(\frac{1}{2}j_{z}^{2}+e^{2}-\frac{5}{2}e_{z}^{2}-\frac{1}{6}\right),\label{eq:phi_quad}
\end{equation}
\begin{equation}
  \phi_{\text{oct}}=\frac{75}{64}\left(e_{x}\left(\frac{1}{5}-\frac{8}{5}e^{2}+7e_{z}^{2}-j_{z}^{2}\right)-2e_{z}j_{x}j_{z}\right),\label{eq:phi_oct}
\end{equation}
\begin{equation}
  \Phi_{0}=\frac{Gm_{per}a^{2}}{a_{per}^{3}\left(1-e_{per}^{2}\right)^{\frac{3}{2}}},\,\,\,\,\,\,\,\epsilon_{\text{oct}}=\frac{a}{a_{per}}\frac{e_{per}}{1-e_{per}^{2}} \label{eq:epsilon_oct}
\end{equation}
with $\Phi_{0}$ and $\epsilon_{\text{oct}}$ constant.
Time and its derivatives are expressed in units of the secular timescale
\begin{equation}
 t_{sec}=\sqrt{GMa}/\Phi_0  \label{eq:tsec}
\end{equation}
using $\tau\equiv t/t_{sec}$. 

\paragraph{Slow variables}
When $\epsilon_\text{oct}=0$ (the periodic analytically solved KLCs) the perturbing potential is axisymmetric (with respect to the \textit{z} axis) admitting a constant of the motion, $j_z$, which limits the eccentricity through the constraint $j>j_z$. KLCs are classified by the values of the constants $j_z$ and
\begin{align}
C_{K}&=\frac{4}{3}\phi_{\text{quad}} + \frac{1}{6} - \frac{1}{2}j^2_z\cr
&=e^{2}-\frac{5}{2}e^2_z. \label{eq:CK}
\end{align}
When $C_K<0$, the argument of pericenter of the inner orbit, $\omega$, librates around $\frac{\pi}{2}$ or $-\frac{\pi}{2}$ (\textit{librating} cycles), and when $C_K>0$, it goes through all values $\left[0,2\pi\right]$ (\textit{rotating} cycles).

In the problem we study here with $\epsilon_\text{oct}>0$, $C_K$ and $j_z$ are no longer constant. The three slow variables $j_z,C_K$ and $\Omega_e$ change slowly on a timescale of $\sim t_{sec}/\sqrt{\epsilon_\text{oct}}$ (see Eqs. \ref{eq:epsilon_oct}-\ref{eq:tsec}) \citep{antognini2015}. As in \cite{katz2011} we focus on the regime of high eccentricity, i.e $\left|j_z\right|\ll1$.

\paragraph{Simple Pendulum}
Up to the leading order in $j_z$ and $\epsilon_\text{oct}$ and averaging over rotating KLCs (librating KLCs with $\left|j_z\right|\ll1$ accumulate change in $j_z$ on a much longer timescale \cite{katz2011}), the averaged equations for the long term evolution of $\Omega_e$ and $j_z$ are \cite{katz2011}
\begin{equation}
\begin{aligned}
  \dot{\Omega}_e & = \left<f_\Omega\right> j_z, \quad & \dot{\phi} & = \frac{1}{L} v \\
  \dot{j}_z & = -\epsilon_\text{oct} \left<f_j\right> \sin \Omega_e, \quad & \dot{v} & = -g \sin \phi
\end{aligned}
\label{eq:averaged_equations}
\end{equation}
where 
\begin{equation}
\begin{aligned}
  \left\langle f_{\Omega}\right\rangle & =\frac{6E\left(x\right)-3K\left(x\right)}{4K\left(x\right)}, \\
  \left\langle f_{j}\right\rangle & =\frac{15\pi}{128\sqrt{10}}\frac{1}{K\left(x\right)}\left(4-11C_{K}\right)\sqrt{6+4C_{K}},\\
  x & =3\frac{1-C_{K}}{3+2C_{K}},
\end{aligned}\label{eq:averaged_functions}
\end{equation}
and $K\left(m\right)$ and $E\left(m\right)$ are the complete elliptic functions of the first and second kind, respectively. For comparison, the equations of motion of a simple pendulum with angle $\phi$ and velocity $v$ along with constants $g$ and $L$ are provided on the right side of Equations \ref{eq:averaged_equations}. Since $\Phi_{\text{per}}$ is constant and $\phi_{\text{oct}}$ and $j_z$ are small - $C_K$ and therefore $\left\langle f_{\Omega}\right\rangle,\left\langle f_{j}\right\rangle$ are approximately constant (see discussion of exceptions below). By comparing the structures of the two sides of Equations \ref{eq:averaged_equations} we see that the averaged equations for $\Omega_e$ and $j_z$ are equivalent to those of a simple pendulum.

The precise correspondence to a simple pendulum depends on the sign of the product $\left\langle f_{\Omega}\right\rangle\left\langle f_{j}\right\rangle$ which in turn depends on the initial conditions through $C^0_K$ having three regions bounded by the zero crossing of $\left<f_\Omega\right>$ at $C^0_K\approx0.112$ and $\left<f_j\right>$ at $C^0_K=\frac{4}{11}$.
The angle $\phi$ of the pendulum is given by the following:\\
For $0.112\lesssim C_{K}^{0}<\frac{4}{11}$
\begin{equation}
  \left<f_{j}\right>\left<f_{\Omega}\right>>0\Rightarrow\phi=\Omega_{e}
\end{equation}
and otherwise 
\begin{equation}
  \left<f_{j}\right>\left<f_{\Omega}\right><0\Rightarrow\phi=\Omega_{e}+\pi.
\end{equation}
The velocity of the pendulum is $\left(+j_z\right)$ for $C^0_K\gtrsim 0.112$ where $\left<f_{\Omega}\right>$ is positive and $\left(-j_z\right)$ otherwise.

\begin{figure}
 \begin{centering}
 \includegraphics[scale=0.25]{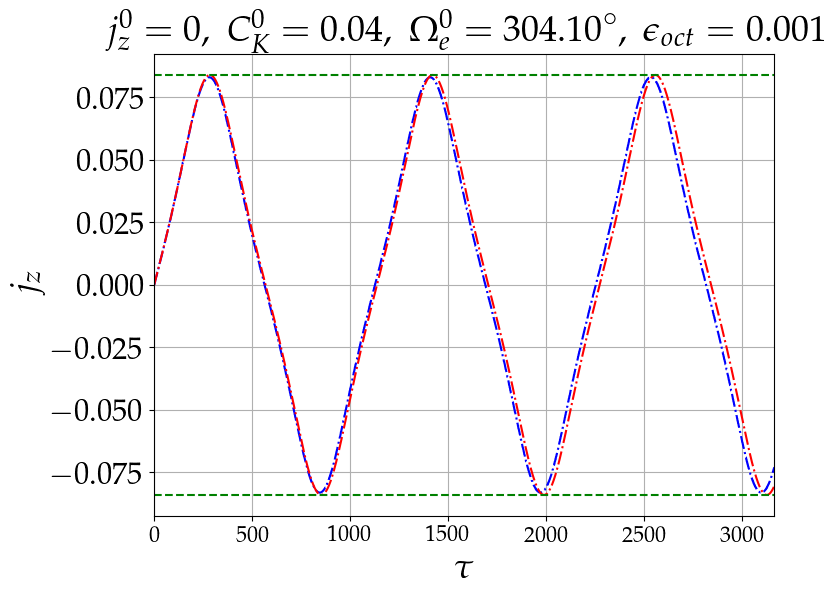}
 \par\end{centering}
 \begin{centering}
 \includegraphics[scale=0.25]{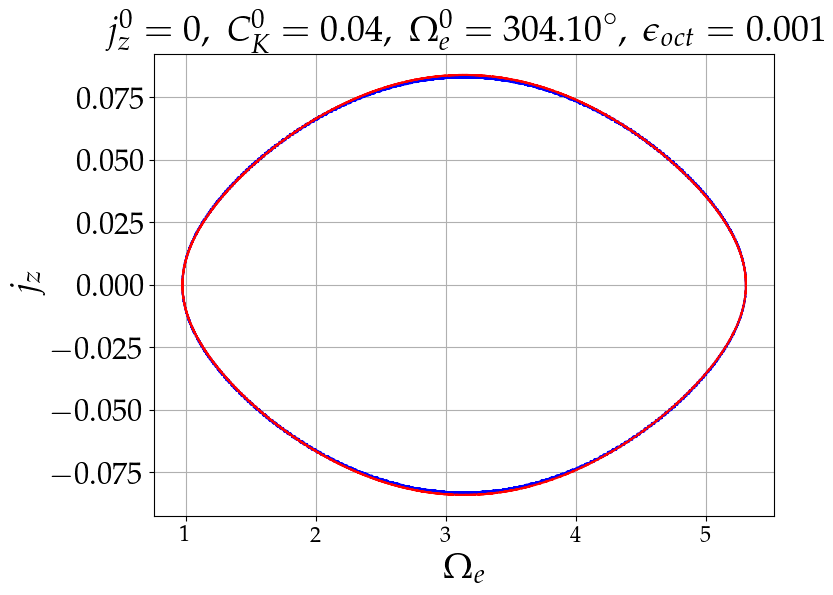}
 \par\end{centering}
 \caption{Results of a numeric integration of the double averaged equations (blue) along with the result of a simple pendulum (Equations \ref{eq:averaged_equations}, red). The values of the initial conditions and $\epsilon_{\text{oct}}$ are shown above the plots. Upper panel: $j_z$ as a function of (normalized) time. The two green horizontal lines are the maximum and minimum values of $j_z$ of the simple pendulum (Equation \ref{eq:jzmax_from_jz0_0}). Lower panel: $j_z$ vs. $\Omega_e$. \label{fig:jz_vs_tau_and_jz_vs_Omega_example}}
\end{figure}

An example of a numerical integration of the full double averaged equations (Equations 4 in \cite{katz2011}, for explicit terms see Equations C3 in \cite{luo2016})) compared with the solution of Equations \ref{eq:averaged_equations} with the approximation of constants $\left\langle f_{\Omega}\right\rangle,\left\langle f_{j}\right\rangle$ (evaluated at $C^0_K$) is shown in Figure \ref{fig:jz_vs_tau_and_jz_vs_Omega_example}. As can be seen - for the example shown - the long term evolution of $j_z$ is successfully approximated by equations of a simple pendulum with velocity $\left(-j_z\right)$.

\paragraph{flip criterion}
A flip - zero crossing of $j_z$ - is equivalent to the velocity of the pendulum changing sign which occurs only if the pendulum is librating.
Given $\epsilon_{\text{oct}}$ and initial values $C^0_K,\Omega^0_e$ the maximal value of $j^0_z$ where a flip occurs is given by
\begin{equation}
j^0_{z,\text{max}}\left(C^0_K,\Omega^0_e\right)=\sqrt{2\epsilon_{\text{oct}}\left|\frac{\left\langle f_{j}\right\rangle }{\left\langle f_{\Omega}\right\rangle }\right|\left(1\pm \cos\Omega_{e}^{0}\right)}\label{eq:jzmax_from_jz0_0}
\end{equation}
where the $\pm$ sign is positive if $\left\langle f_{j}\right\rangle \left\langle f_{\Omega}\right\rangle > 0$ and negative otherwise. A global flip criterion is thus 
\begin{equation}
j^0_{z,\text{max}}\left(C^0_K\right)=2\sqrt{\epsilon_{\text{oct}}\left|\frac{\left\langle f_{j}\right\rangle }{\left\langle f_{\Omega}\right\rangle }\right|}\label{eq:jzmax_max_from_jz0_0}
\end{equation}
or equivalently
\begin{equation}
\epsilon_{\text{oct}}>\frac{1}{4}\left|\frac{\left\langle f_{\Omega}\right\rangle }{\left\langle f_{j}\right\rangle }\right|\left(j_{z}^{0}\right)^{2}.\label{eq:flip_criterion_for_epsilon_oct}
\end{equation}

\begin{figure}
 \begin{centering}
 \includegraphics[scale=0.4]{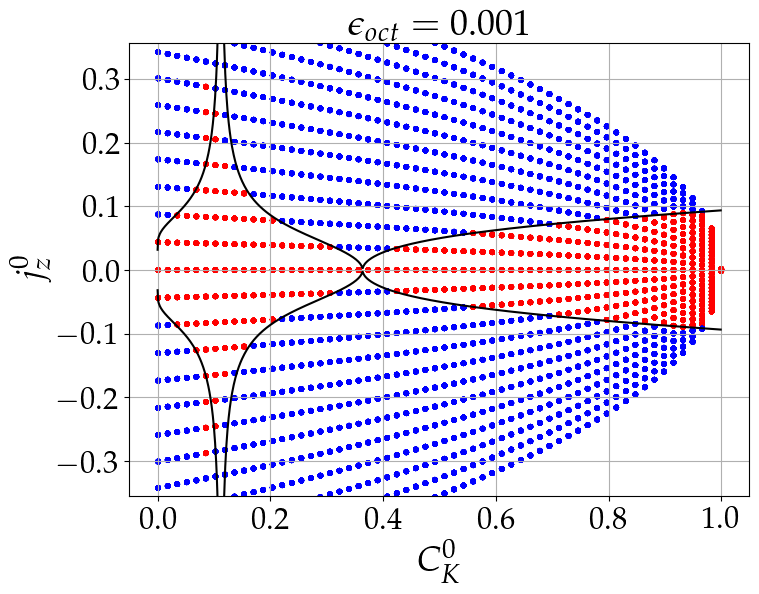}
 \par\end{centering}
 \caption{Parameter space of orbital flips from the solution of the double averaged secular equations (up to $\tau=10/\epsilon_\text{oct}$). Each point represents $36$ numerical integrations with $\omega^0=0$ and different $\Omega^0$ scanned between $0$ and $2\pi$. Red points mark integrations where $j_z$ crossed zero during the evolution for some $\Omega^0$ and blue points mark those where $j_z$ kept its sign for all $\Omega^0$. The black line is the analytic maximal value of $j^0_z$ allowing a flip obtained by the simple pendulum model (Equations \ref{eq:jzmax_max_from_jz0_0} and \ref{eq:averaged_functions}).\label{fig:flipmap}}
\end{figure}

A comparison between a numerical \textit{flip map} and the analytic prediction of Equations \ref{eq:jzmax_max_from_jz0_0} and \ref{eq:averaged_functions} as a function of $j^0_z$ and $C^0_K$ is shown in Figure \ref{fig:flipmap} for $\epsilon_{\text{oct}}=0.001$. Each point represents a set of numerical simulations (up to $\tau=10/\epsilon_\text{oct}$) with the same $j^0_z$ and $C^0_K$, $\omega^0=0$ and a range of values for the longitude of ascending node, $\Omega^0$ (which is equal to $\Omega_e$ when $\omega=0$) equally spaced by $10^\circ$ between $0$ and $360^\circ$. The values of $e^0$ and $i^0$ are set by the choice of the other parameters. A point is marked red if for some $\Omega^0$ a flip occurred and blue otherwise. A black line marks the analytic prediction of Equations \ref{eq:jzmax_max_from_jz0_0} and \ref{eq:averaged_functions}. As can be seen, the black line follows the border between red and blue points with some exceptions discussed below.

\begin{figure}
 \begin{centering}
 \includegraphics[scale=0.4]{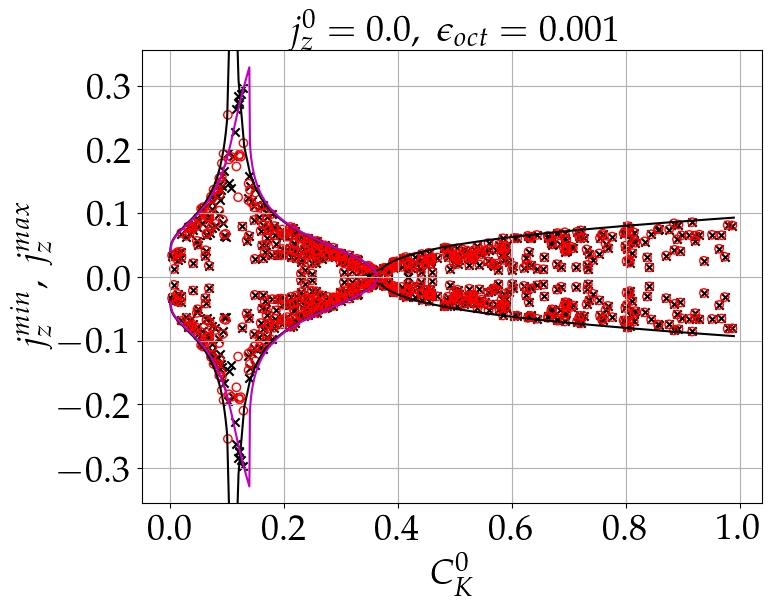}
 \par\end{centering}
 \caption{$j^{\text{max}}_z$ and $j^{\text{min}}_z$ vs. $C^0_K$ when $j^0_z=0$ and all other components randomly chosen (uniformly distributed in $e_x,e_y$ and $j_x$). The results of direct numerical integrations of the double averaged equations (up to $\tau=10/\epsilon_\text{oct}$) are shown using black crosses. The analytic results of the simple pendulum model are marked with open red circles (through Equation \ref{eq:jzmax_from_jz0_0}) and a black line (envelope, Equations \ref{eq:jzmax_max_from_jz0_0} and \ref{eq:averaged_functions}). The expected envelope in the (more complicated) analytic model of \cite{katz2011} (using Equations 14 and 15 therein) is shown with a magenta line for $C_K<\frac{4}{11}$.\label{fig:jz_minmax_vs_CK_numerical_all_points}}
\end{figure}

The maximal and minimal values of $j_z$ obtained in $391$ numerical simulations with $j^0_z=0$ and randomly chosen initial conditions (uniformly distributed in $e_x,e_y$ and $j_x$) is plotted using black crosses in Figure \ref{fig:jz_minmax_vs_CK_numerical_all_points} for $\epsilon_{\text{oct}}=0.001$. The maximal deviation from $j^0_z=0$, $j^\text{max}_z$, represents the maximal $j^0_z$ that lead to a flip assuming $C_K$ is constant (see below discussion of changes in $C_K$). The black line in Figure \ref{fig:jz_minmax_vs_CK_numerical_all_points} comes from Equations \ref{eq:jzmax_max_from_jz0_0} and \ref{eq:averaged_functions} and is identical to the black line in Figure \ref{fig:flipmap}. As can be seen, the black line agrees with the envelope of the numerically available $j^\text{max}_z$. The open circles in Figure \ref{fig:jz_minmax_vs_CK_numerical_all_points} denote the analytical predicted values using Equations \ref{eq:jzmax_from_jz0_0} and \ref{eq:averaged_functions}\footnote{In contrast to the maximal value of $j^0_z$ allowing a flip, for the maximal and minimal values of $j_z$ starting from $j^0_z=0$ the $\pm$ sign of Equation \ref{eq:jzmax_from_jz0_0} is positive if $\left\langle f_{j}\right\rangle \left\langle f_{\Omega}\right\rangle < 0$ and negative otherwise}. As can be seen, the predictions of Equation \ref{eq:jzmax_from_jz0_0} for different values of $\Omega^0_e$ agree with the numerical results to an excellent approximation.

As can be seen in Equation \ref{eq:jzmax_max_from_jz0_0} and in the black lines in Figures \ref{fig:flipmap}-\ref{fig:jz_minmax_vs_CK_numerical_all_points} at $C^0_K\approx0.112$ where $\left\langle f_{\Omega}\right\rangle=0$, $j^\text{max}_z$ diverges. At this point, $\Omega_e$ is constant so the accumulation in $j_z$ continues indefinitely \cite{katz2011}. The numerical results do have a significant increase in that vicinity of $C^0_K$ but the divergence is saturated and some points deviate from the analytic prediction. A saturation is expected when taking into account the small change in $C_K$ during the evolution allowing $\left\langle f_{\Omega}\right\rangle$ to deviate from zero. The maximal (and minimal) values of $j_z$ starting at $j^0_z=0$ based on Equations 14 and 15 from \cite{katz2011} is shown in Figure \ref{fig:jz_minmax_vs_CK_numerical_all_points} using a magenta line. As can be seen, the magenta line does not diverge and agrees with the location and value of the maximal saturation of $j_z$. Note that when $C_K$ changes, the location of this maximum does not correspond to the location in the flip map and requires a (similar) separate criterion.

\begin{figure}
 \begin{centering}
 \includegraphics[scale=0.4]{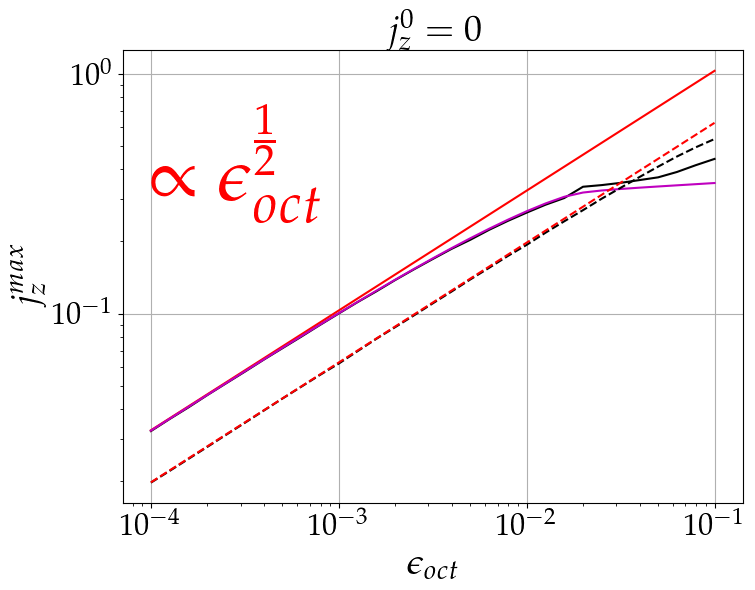}
 \par\end{centering}
 \caption{The maximal attainable value of $j^{\text{max}}_z$ vs. $\epsilon_{\text{oct}}$ for initial conditions with $j^0_z=0$ for two values of $C^0_K$ ($C^0_K=0.05$ solid lines and $C^0_K=0.6$ dashed lines). Shown are results from a numerical solution of the double averaged equations using black lines. Red lines mark the prediction of the simple pendulum model (Equations \ref{eq:jzmax_max_from_jz0_0} and \ref{eq:averaged_functions}, $\propto \epsilon^\frac{1}{2}_{\text{oct}}$). For $C^0_K=0.05$ the analytic prediction of \cite{katz2011} (using Equations 14 and 15 therein) is plotted in magenta.\label{fig:jz_max_vs_eqpsilon}}
\end{figure}

An additional region of initial conditions where $\left\langle f_{\Omega}\right\rangle$ and $\left\langle f_{j}\right\rangle$ can significantly change due to the small change in $C_K$ is around $C^0_K\ll1$ which is the focus of the analysis of \cite{katz2011}. In this region the small change in $C_K$ can be of the order of $C^0_K$ itself. In Figures \ref{fig:flipmap} and \ref{fig:jz_minmax_vs_CK_numerical_all_points} where $\epsilon_\text{oct}=0.001$ the deviation is not apparent. In Figure \ref{fig:jz_max_vs_eqpsilon} the dependence of the maximal available value of $j^{\text{max}}_z$ (starting from $j^0_z=0$) on $\epsilon_{\text{oct}}$ is presented for low and high values of $C^0_K$, $C^0_K=0.05$ and $C^0_K=0.6$ respectively. As can be seen, at $C^0_K=0.05$ and for high values of $\epsilon_\text{oct}$ the numerical results (solid black line) deviate from the simple pendulum model (solid red line, $\propto \epsilon^\frac{1}{2}_\text{oct}$). The prediction of \cite{katz2011} (solid magenta) agrees with the numerical results to much higher values of $\epsilon_\text{oct}$. In contrast, for the high value of $C^0_K=0.6$ shown (dashed lines) the numerical results follow the simple pendulum dependence across a wide range of $\epsilon_\text{oct}$ values.

\paragraph{Discussion}

The simple pendulum model is valid for most initial conditions because $C_K$, $\left\langle f_{\Omega}\right\rangle$ and $\left\langle f_{j}\right\rangle$ are constant for most values of $C^0_K$ (see Equation \ref{eq:CK} and discussion below). There are two small regions of phase space where the approximation of constant $C_K$ fails to reconstruct the dynamics: (i) At the vicinity of $C_K\approx0.112$ where $\left\langle f_{\Omega}\right\rangle$ approaches zero the simple pendulum model predicts diverging change in $j_z$.
(ii) For $C^0_K\ll1$ where $C_K$ can change by the order of itself. As shown, the more complicated analytic solution constructed in \cite{katz2011} manages to approximately take into account the change in $C_K$ in these cases.

Examining the derivation of the change in \(C_K\) presented in \cite{katz2011} reveals a logical error that appears to contradict the successful agreement with numerical results. The change in $C_K$ is approximated to be equal to the change in $\left(-\frac{1}{2}j^2_z\right)$ based on approximating $\phi_\text{quad}$ as constant (in Equation \ref{eq:CK}). This is not justified since $\phi_\text{quad}$ is constant only up to $\sim\epsilon_\text{oct}$ while $j^2_z$ is comparable or smaller. A closer inspection reveals the reason why this error does not affect the results. For low enough values of $e_{\text{min}}$ (and $\left|j_z\right|\ll1$) - applicable for the two regions where $\left\langle f_{\Omega}\right\rangle$ and $\left\langle f_{j}\right\rangle$ change - $\phi_{\text{oct}}$ (Equation \ref{eq:phi_oct}) is much smaller than $1$ throughout most of the time during each KLC. The small time spent at high $e$ where $\phi_{\text{oct}}\sim 1$ is not sufficient for accumulation of a significant change in $j_z$ and $\Omega_e$. In contrast, for regions of $e_{\text{min}}\sim1$ ($C^0_K\sim1$), the change in $C_K$ is significantly different from the change in $\left(-\frac{1}{2}j^2_z\right)$ but in these regions it is not important since the functions $\left\langle f_{\Omega}\right\rangle,\left\langle f_{j}\right\rangle$ do not change significantly.

Finally, we note that beyond a simpler and more intuitive analytical model for EKL, the approach presented in this Letter allows an extension \citep{Klein2024_CDA} to the analytic solution that includes corrections to the double averaging that are important in the presence of a massive perturber (e.g \cite{luo2016,tremaine2023}).

We thank Smadar Naoz, Scott Tremaine and Chris Hamilton for useful discussions.

\bibliography{octupole_is_a_pendulum}

\end{document}